\documentclass[aps,pre,
reprint,
 notitlepage,
 showpacs,floatfix,nofootinbib,
 superscriptaddress
 ]{revtex4-1}

 \usepackage{subfigure}
 \usepackage{amssymb}
 \usepackage{amsfonts}
 \usepackage{amsmath}
 \usepackage{amsthm}
 \usepackage{epsfig}
 \usepackage{float}
 \usepackage[usenames,dvipsnames]{color}
 \usepackage[latin1]{inputenc}

 \usepackage{graphics,graphicx,xcolor,subfigure}

 \usepackage[hidelinks,unicode=true]{hyperref}
 \hypersetup{colorlinks=true,
 	linkcolor=blue,
 	urlcolor=blue,
 	citecolor=blue,
 	pdfhighlight=/N
 }
 
 \usepackage{comment}
 \usepackage[normalem]{ulem}
 \usepackage{microtype}

\newcommand{\av}[1]{\langle {#1} \rangle}

\newcommand{\Ninf}{N_\mathrm{inf}}

\newcommand{\NSI}{N_\mathrm{SI}}
\newcommand{\kmax}{k_\text{max}}
\newcommand{\kmin}{k_\text{min}}

\newcommand{\aqmf}{a_{_\text{QMF}}}
\newcommand{\apqmf}{a_{_\text{PQMF}}}
\newcommand{\knn}{\kappa_\text{nn}}
\newcommand{\ave}[1]{\langle {#1} \rangle_\text{e}}

\begin{document}

\title{High prevalence regimes in the pair quenched mean-field theory 
	for the susceptible-infected-susceptible model on networks}

\author{Diogo H. Silva}
\affiliation{Departamento de F\'{\i}sica, Universidade Federal de Vi\c{c}osa, 36570-900 Vi\c{c}osa, Minas Gerais, Brazil}

\author{Francisco A. Rodrigues}
\affiliation{Instituto de Ci\^{e}ncias Matem\'{a}ticas e de Computa\c{c}\~{a}o, Universidade de S\~{a}o Paulo, S\~{a}o Carlos, SP, Brazil.}

\author{Silvio C. Ferreira}
\affiliation{Departamento de F\'{\i}sica, Universidade Federal de Vi\c{c}osa, 36570-900 Vi\c{c}osa, Minas Gerais, Brazil}
\affiliation{National Institute of Science and Technology for Complex Systems, 22290-180, Rio de Janeiro, Brazil}

\begin{abstract} 
Reckoning of pairwise dynamical correlations significantly   improves the
accuracy of mean-field theories and plays an important role in the investigation
of dynamical processes on complex networks. In this work, we perform a
nonperturbative numerical analysis of the quenched mean-field theory (QMF) and
the  inclusion of dynamical correlations by means of the pair quenched mean-field (PQMF) theory
for the susceptible-infected-susceptible (SIS) model on synthetic and real
networks.  We show that the PQMF considerably outperforms the standard QMF on
synthetic networks of distinct levels of heterogeneity and degree correlations,
providing extremely accurate predictions when the system is not too close to the
epidemic threshold while the QMF theory deviates substantially from simulations
for networks with a degree exponent $\gamma>2.5$. The scenario for real networks
is more complicated, still with PQMF significantly outperforming   the QMF
theory. However, despite of high accuracy for most investigated networks, in a
few cases PQMF deviations from simulations are not negligible. We found 
correlations between accuracy and average shortest path while other basic
networks metrics seem to be uncorrelated with the theory accuracy. Our results
show the viability of the PQMF theory to investigate the high prevalence regimes
of recurrent-state epidemic processes on networks, a regime of high
applicability.
\end{abstract}


\maketitle

\section{Introduction}
Pairwise approximation  constitutes a valuable tool recurrently used for
understanding dynamical processes on graphs (networks or lattices) and,
particularly, epidemic spreading on the top of complex
networks~\cite{kiss2017mathematics,Wang2017,DeArruda2018}. This approach
outperforms ordinary mean-field approximations extending dynamical equations
from one-site to a pair level~\cite{Marro2005}. Extensions to higher orders
methods using $n$-cluster approximations~\cite{Ben-Avraham1992} can lead to more
accurate theories at the cost of increasing  the theoretical complexity. While
being of limited application for low dimensional systems near to critical phase
transitions~\cite{Marro2005}, pair approximations can be remarkable improvements
with respect to the one-site theory if either we are not too close to the
transition~\cite{Joo2004} or if the system dimension is large such as the case of
random graphs~\cite{Gleeson2011,Gleeson2013}.

For dynamical processes on the top of complex networks, heterogeneities play a
central role~\cite{Barrat2008,PastorSatorras2015} that has to be taken into
account to reproduce the most fundamental results~\cite{Pastor-Satorras2001,
	YangWang}. Particularly, the interplay between structural heterogeneity and 
dynamical correlations has been investigated using heterogeneous
pair-approximations~\cite{Eames2002, Keeling1999, Gleeson2011, Gleeson2013,
	Mata2014}. We consider the susceptible-infected-susceptible (SIS)
model~\cite{Barrat2008} whose dependence on heterogeneities serves as reference
for many other dynamical processes~\cite{Barrat2008,Rodrigues2016}. In the SIS
model individuals are represented by vertices of a network and can be in either
susceptible or infected states. Infected vertices heals spontaneously with rate
$\mu$ and infect their susceptible neighbors with rate $\lambda$ per contact.  A
central  aspect of spreading phenomena is the epidemic threshold
$\lambda_\text{c}$ above which an extensive fraction  of the population is
infected or, in other words, the epidemic prevalence is finite.

Heterogeneities of networks can change drastically the behavior of the
threshold. If the network possesses a heavy-tailed degree distribution in the
form of a power-law $P(k)\sim k^{-\gamma}$ the epidemic threshold of SIS model
is zero in the thermodynamical limit when the network size goes to
infinity~\cite{Chatterjee2009, Mountford2016, Castellano2010}. This involves  a
very special type of transition from an active and fluctuating  to an absorbing
state~\cite{Castellano2012,Boguna2013} which can be knocked out with small
modifications of the SIS dynamics~\cite{Ferreira2016a,Cota2018a}. The SIS transition
on uncorrelated power-law networks can be of two types~\cite{Castellano2012,
	Boguna2013, Ferreira2016a}: if the degree exponent is small ($\gamma<2.5$) 
the activation is triggered by a densely connected core identified by the
maximal index of a $k$-core decomposition~\cite{Castellano2012}. If the degree
exponent is large ($\gamma>2.5$) then the activation is ruled by the hubs. 
{The latter involves long-term epidemic activity on star subgraphs, composed of a
single hub (the center) and its $k_\text{hub}$ nearest-neighbors (the leaves),
through a feedback mechanism where the hub infects the leaves which in turn
reinfect the hub.  This activity must last for sufficiently long  times to
permit the mutual activation of hubs which are not directly connected
(long-range) \cite{Chatterjee2009, Mountford2016, Boguna2013}.}

Heterogeneities can be included in mean-field approximations in different
forms~\cite{PastorSatorras2015}. Two widely used approximations are the
heterogeneous mean-field (HMF)~\cite{Pastor-Satorras2001} and quenched
mean-field (QMF)~\cite{YangWang,Chakrabarti2008} theories. The former consists of a
cross-graining where only the degree of the nodes and the statistical degree
correlations are  included in the dynamical equations for the probability that a node
is infected~\cite{Pastor-Satorras2001,Boguna2002}, and neglects the dynamical
correlations. The latter includes the full connectivity structure of the
networks but still neglects dynamical correlations assuming that the states of
nearest-neighbors are independent. Due to the aforementioned nature of the SIS
activation mechanisms, the explicit inclusion of the network connectivity
structure, as in the QMF approach, is imperative to construct mean-field
theories of the SIS model since heterogeneous mixing, which corresponds to an
annealed network~\cite{Castellano2012}, will destroy localization effects as for example the self-sustained activity in a star subgraph.

Despite of the detailed microscopic description of the QMF theory, neglecting
dynamical correlations in SIS model can lead to modest accuracy with significant
deviations from simulations~\cite{Ferreira2012,Mata2013,Silva2019} if epidemic
involves, for example, activation localized in the hubs that spreads to the rest
of network~\cite{Chatterjee2009,Castellano2012}. Indeed, the threshold predicted
by the QMF theory for the SIS model, given by the inverse of the largest
eigenvalue of the adjacency matrix~\cite{YangWang,Castellano2010} (see
Sec.~\ref{sec:qmf} for details), involves a localized phase on random power-law
networks with degree exponent $\gamma>2.5$~\cite{Goltsev2012}. Dynamical
correlations reckoned by individual pairwise interactions greatly improve the
predictions of the epidemic thresholds of the QMF approach in  the so-called pair QMF
(PQMF) theory~\cite{Mata2013}; see Sec.~\ref{sec:qmf} for details. Indeed, PQMF
theory~\cite{DeArruda2018,Wang2017,Silva2019,St-Onge2017} and modified versions
of it~\cite{Matamalas2018,Wu2018,Wu2019,Wu2020} have been intensively investigated
recently. The asymptotic scaling  exponent of the threshold as function of the
network size is unchanged when the pair-approximation is included in the QMF
theory~\cite{Mata2013,Silva2019}. 

Since PQMF theory has been mainly analyzed perturbatively in the limit of very
low prevalence to investigate the position of the epidemic thresholds, Matamalas
\textit{et al}.~\cite{Matamalas2018} claimed that it has limitations to compute
high epidemic incidence regimes and proposed that a microscopic Markov chain approach (MMCA)~\cite{Gomez2010}, which is a discrete time version of the QMF theory, could be used instead. However, a nonperturbative approach is possible though numerical integration of
both QMF and PQMF dynamical equations. Since large discrepancies between
discrete and continuous time approaches can be present in the SIS
dynamics~\cite{Fennell2016a}, a nonperturbative analysis of QMF and PQMF
theories is necessary. We develop nonperturbative analyses of both QMF and
PQMF theories for SIS on networks using numerical integration of the
corresponding dynamical equations. We consider both large synthetic networks
generated with the Weber-Porto model~\cite{Weber2007} and real networks with
different levels of degree correlation. We address regimes not asymptotically
close to the epidemic threshold since the mean-field theories fail in
predicting  very low densities of infected
vertices~\cite{Silva2019,Mountford2013}. However, numerical analyses supports
that this asymptotic scaling is confined into a small interval near to
$\lambda=\lambda_\text{c}\rightarrow 0$ such that the mean-field theories are
still applicable beyond this region; see Fig.~\ref{fig:g230a0}. We report that
the PQMF theory predicts very accurately the epidemic prevalence in synthetic
networks for all ranges of degree exponent ($\gamma<2.5$ and $\gamma>2.5$) and
correlations (assortative, disassortative, or uncorrelated) investigated while
QMF theory presents significant deviations for $\gamma>2.5$. For real networks,
in general, PQMF theory considerably outperforms QMF  but also presents
non-negligible deviations from simulations in some cases.

The remaining  of the paper is organized as follows. The mean-field theories used in 
this work are discussed in Sec.~\ref{sec:qmf}. The comparison of numerical integration
and stochastic simulations are performed in Sec.~\ref{sec:result} while our concluding
remarks are presented in Sec.~\ref{sec:conclu}. Appendices \ref{app:wp}, \ref{app:qs}, and \ref{app:algo}
present technical details of our numerical analyses.

\section{Mean-field theories}
\label{sec:qmf}

We investigate the SIS model on a connected, undirected, and unweighted network with
$i=1,\ldots,N$  vertices whose structure is encoded in the adjacency matrix $A_{ij}$ defined by
$A_{ij}=1$ if $i$ and $j$ are connected and $A_{ij}=0$ otherwise. The healing rate is fixed
to $\mu=1$ without loss of generality.

The probability that a vertex $i$ is infected, represented
by $\rho_i$, evolves as~\cite{Mata2013}
\begin{equation}
\frac{d\rho_i}{dt} = -\rho_i+\lambda\sum_j\phi_{ij}A_{ij}
\label{eq:rhoi_pair}
\end{equation}
where $\phi_{ij}$ is the probability that a vertex $i$ is susceptible and its
nearest-neighbor $j$ is infected. Equation~\eqref{eq:rhoi_pair} is exact but not
closed. A closed system is obtained taking the one-site approximation
$\phi_{ij}\approx\rho_i(1-\rho_j)$ that correspond to the QMF
theory~\cite{Chakrabarti2008,Goltsev2012}
\begin{equation}
\frac{d\rho_i}{d t} = -\rho_i+\lambda(1-\rho_i)\sum_{j=1}^{N}A_{ij}\rho_j.
\label{eq:qmf}
\end{equation}
The QMF epidemic threshold is given by $\lambda_\text{c}^\text{QMF} \Lambda^{(1)}=1$
where $\Lambda^{(1)}$ is the largest eigenvalue (LEV) of the adjacency matrix $A_{ij}$.

The PQMF theory includes dynamical correlations considering the evolution of
$\phi_{ij}$ which, in its complete but not closed form, depends on the triplets;
see Ref.~\cite{Mata2013}. The PQMF theory consists of approximating the triplets
$[A_i,B_j,C_l]$ in which  $i$ and $l$ are both connected to $j$  by
\begin{equation}
[A_i,B_j,C_l]\approx \frac{[A_i,B_j][B_j,C_l]}{[B_j]}.
\label{eq:pair}
\end{equation} 
Here $A$, $B$, and $C$ are the states of the vertices that, in the SIS case, can
be either infected or susceptible. The approximation given by
Eq.~\eqref{eq:pair} considers that $(i,j,l)$ does not form a triangle,
i.e., $i$ and $l$ are connected to $i$ but not to each other. Actually, the
effects of clustering have been recently investigated~\cite{Wu2020} and it was
shown that even in networks with high cluster coefficient, plenty of
triangles, the approximation given by Eq.~\eqref{eq:pair}  performs very well for
the steady state.  The final PQMF equation for $\phi_{ij}$
becomes
\begin{eqnarray}
\frac{d\phi_{ij}}{dt}   & = &  -(2+\lambda)\phi_{ij}+\rho_j
+\lambda\sum_{l} \frac{\omega_{ij}\phi_{jl}}{1-\rho_j}(A_{jl}-\delta_{il}) \nonumber \\
& - & \lambda\sum_{l} \frac{\phi_{ij}{\phi}_{il}}{1-\rho_i}(A_{il}-\delta_{lj}) ,
\label{eq:phi2}
\end{eqnarray}
in which $\omega_{ij}=1-\phi_{ij}-\rho_i$. Equations~\eqref{eq:rhoi_pair} and \eqref{eq:phi2} form a
closed system of $N+M$ equations where $M=\frac{1}{2}\sum_j k_i$ is the number of 
edges of the network. More details of the derivation are given in Ref.~\cite{Mata2013}.

The epidemic threshold within the PQMF framework is given by the transcendent equation $\lambda_\text{c} \Omega^{(1)}(\lambda_\text{c})=1,$~\cite{Silva2019} where $\Omega^{(1)}$ is the largest eigenvalue of the weighted adjacency matrix  $B_{ij}$ given by
\begin{equation}
B_{ij}=\frac{2+\lambda}{2\lambda+2}\frac{A_{ij}}{1+\frac{\lambda^{2}k_{i}}{2\lambda+2}}.
\label{eq:Bij}
\end{equation}
See~\cite{Silva2019} for the derivation of Eq.~\eqref{eq:Bij}.

Very close to the  threshold, the epidemic prevalence $\rho$, defined
as $\rho=\frac{1}{N}\sum_i\rho_i$, approaches zero  following a power-law in the
form $\rho\simeq a_1 (\lambda-\lambda_c)^\beta$ where $\beta$ is a critical
exponent~\cite{Marro2005} and $a_1$ is prefactor that may depend on the network
size. Either in QMF~\cite{Goltsev2012,VanMieghem2011} and PQMF~\cite{Silva2019}
theories it can be shown that $\beta_\text{QMF}=\beta_\text{PQMF}=1$ while
$\aqmf$ or $\apqmf$ can be expressed in terms of the  principal eigenvector (PEV)
$\lbrace v_i^{(1)} \rbrace$ of either $A_{ij}$ or $B_{ij} (\lambda_\text{c})$,
respectively, as~\cite{Silva2019,Goltsev2012}
\begin{equation}
a= \frac{\sum_{i=1}^{N} v^{(1)}_i}{N \sum_{i=1}^{N} \left[v^{(1)}_i\right]^3}.
\label{eq:alpha1}
\end{equation}
These results are straightforwardly derived when the network presents a spectral
gap $\Lambda^{(1)}\gg\Lambda^{(2)}$, where $\Lambda^{(2)}$ is the second LEV of
the adjacency matrix; see e.g.~\cite{Goltsev2012,Silva2019}. However, it was
shown that $\beta_\text{QMF}=1$ is always true~\cite{Mieghem2012} and the same
is expected for the PQMF theory since pair approximations should not change the
universality class predicted by the one-vertex theory~\cite{Ben-Avraham1992}.

\begin{figure}[hbt]
	\centering
	\includegraphics[width=0.9\linewidth]{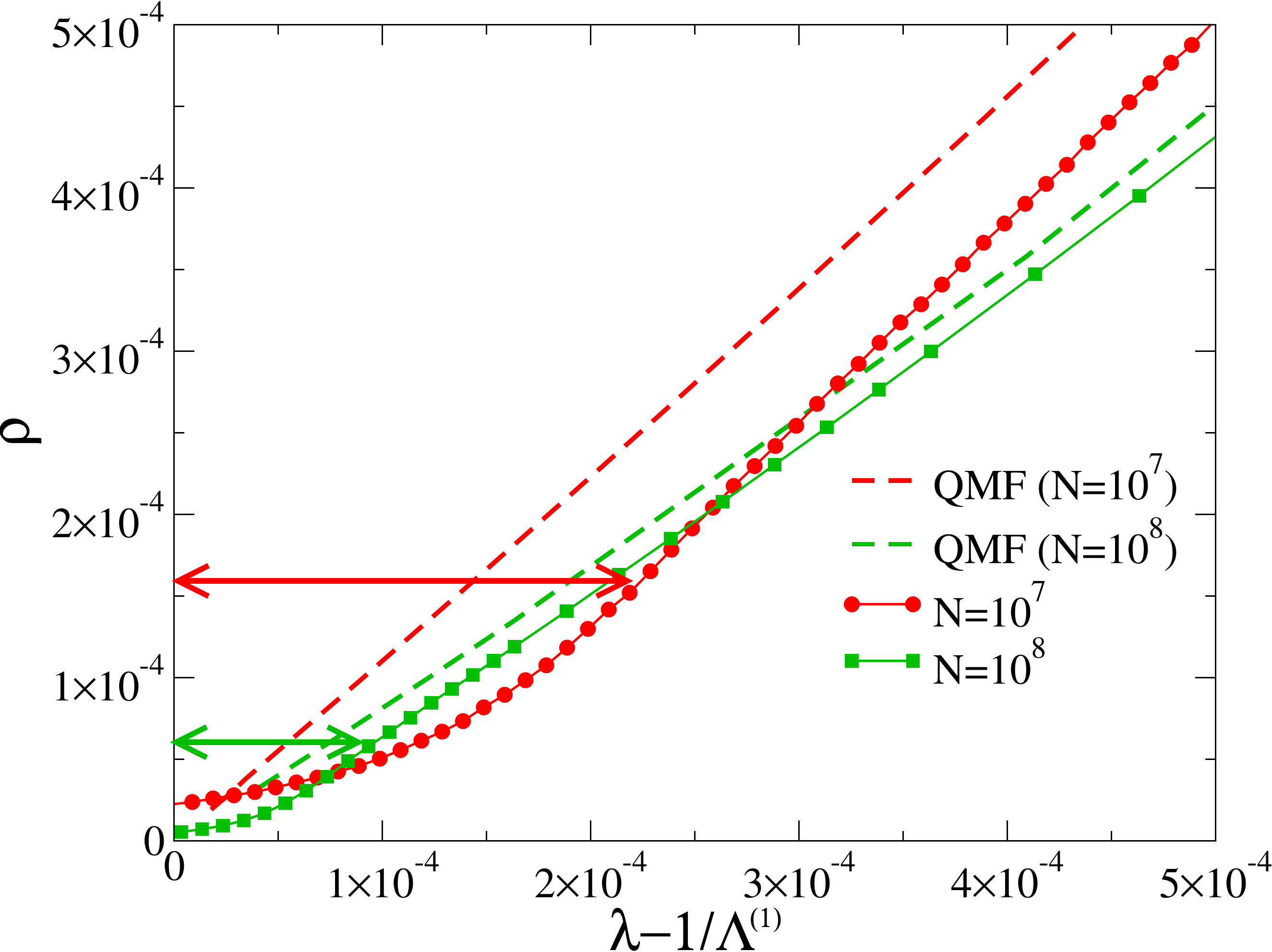}
	\caption{Epidemic prevalence around the transition threshold
		$\lambda_\text{c}=1/\Lambda^{(1)}$ for UCM networks with $\gamma=2.3$,
		$\kmin=3$, and $\kmax=2\sqrt{N}$. Simulations are represented by solid lines
		with symbols and numerical integration of the QMF equation~\eqref{eq:qmf} is
		given by the dashed lines. Horizontal arrows indicate the interval where curves
		depart from linearity.}
	\label{fig:g230a0}
\end{figure}
The mean-field exponent $\beta=1$ does not match the rigorous results obtained by Mountford \textit{et al}.~\cite{Mountford2013} in thermodynamical limit $N\rightarrow\infty$ where
\begin{equation}
\rho \sim  \left\lbrace  \begin{array}{l l l}
 \lambda^{\frac{1}{3-\gamma}} & \text{if} & 2<\gamma<5/2 \\
 \frac{\lambda^{2\gamma -3}}{(\ln \frac{1}{\lambda})^{\gamma-2}} & \text{if}& 5/2<\gamma<3 \\
  \frac{\lambda^{2\gamma-3}}{(\ln \frac{1}{\lambda})^{2\gamma-4}} & \text{if}& \gamma>3,
\end{array} \right.
\label{eq:scaMount}
\end{equation}
according to which $\beta>1$ for any $\gamma>2$.  For large networks with
$\gamma<5/2$, where the epidemic threshold is very accurately reproduced by the QMF
theory, the numerical integration performed in
Ref.~\cite{Silva2019}  confirms the deviation  from the exact result for
$\lambda$ approaching $\lambda_\text{c}^\text{QMF}=\frac{1}{\Lambda^{(1)}}$
while stochastic simulations are in agreement with the rigorous results.
However, the simulations show a pre-asymptotic behavior fully consistent  with
$\beta_\text{QMF}=1$. In a linear scale\footnote{Obviously, this region will not
	shrunk in a logarithm scale and the asymptotic scaling is the theoretical one
	given by Eq.~\eqref{eq:scaMount}.}, the region that departs from linearity is
squeezed around $\lambda=\lambda_\text{c}\rightarrow 0$ as the network size
increases, as indicated by the horizontal arrows in Fig.~\ref{fig:g230a0}, in
which simulations and QMF theory are compared. The slope of the linear region
decreases with size since $\aqmf$ also does: we found $\aqmf=0.00382$  for $N=10^7$
and $\aqmf= 0.00130$ for $N=10^8$. Finally, we see that QMF is not able to
capture quantitatively the amplitude of the linear region observed in
simulations reinforcing the need of nonperturbative analyses of the PQMF
theory.

A closed solution for Eqs~\eqref{eq:rhoi_pair} and \eqref{eq:phi2} can be
derived for the particular case of homogeneous networks where
$P(k)=\delta_{k,m}$ for which $\rho_i=\rho$ and $\phi_{ij}=\phi$. The expression
for stationary epidemic prevalence is\footnote{The solution is same derived for
	the contact process in, e.g., Ref.~\cite{Ferreira2013} replacing the infection
	rate $\lambda_\text{CP} = m\lambda_\text{SIS}$, where $\lambda_\text{CP}$ is the
	infection rate for the contact process and $\lambda_\text{SIS}$ for SIS.}
\begin{equation}
\bar{\rho} = \frac{\lambda-\lambda_\text{c}}{m^{-1}+\lambda-\lambda_\text{c}},~~~~
\lambda_\text{c} = \frac{1}{m-1}.
\label{eq:phomo}
\end{equation}

\section{Results}
\label{sec:result}

We numerically integrated QMF and PQMF equations using a fourth-order
Runge-Kutta method with time step $\delta t= 10^{-2}$ to $10^{-1}$. Initial conditions consistent with the exact closure equations
relating pairwise and single vertex probabilities such as
$[S_i,I_j]+[I_i,I_j]=[I_j]$ must be chosen and the steady state is insensitive to a particular choice. We performed stochastic simulations of the SIS
dynamics on networks using an optimized Gillespie algorithm~\cite{Cota2017}; see
Appendix~\ref{app:algo}. The absorbing states, which are rigorously the unique
real stationary state in finite size networks,  were circumvented using
quasistationary (QS) simulations~\cite{Sander2016}; see Appendix~\ref{app:qs}.

\subsection{Synthetic networks}
A simple metrics to quantify the correlations is the average degree of the
nearest-neighbors of vertices with a given degree
$k$~\cite{Pastor-Satorras2001b,Barrat2008}, represented by $\knn(k)$. The
functional form of $\knn(k)$ reveals correlation patterns of the network. If
$\knn$ is an increasing function of $k$, the network presents assortative
correlations where vertices of similar degree tend to be connected. Conversely,
if $\knn$ decreases with $k$  the network has disassortative correlations where
vertices of high degree tend to be connected with vertices of low degree. Finally, if $\knn$ does not depend of the degree, the networks is
said uncorrelated or neutral and assumes de value
$\knn=\av{k^2}/\av{k}$~\cite{Barrat2008}.

\begin{figure}[!tbh]
	\centering
	\includegraphics[width=0.8\linewidth]{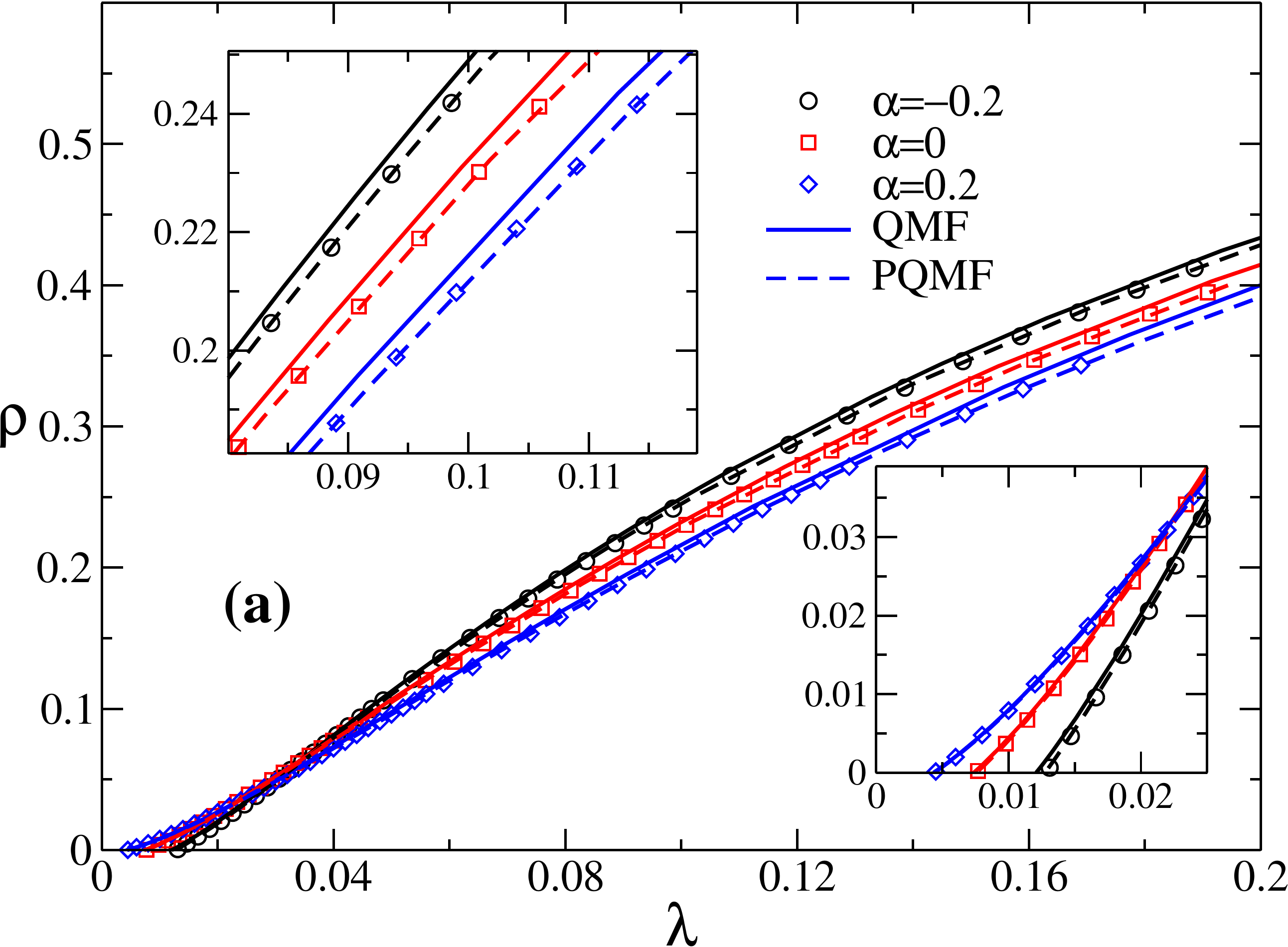}\\
	\includegraphics[width=0.8\linewidth]{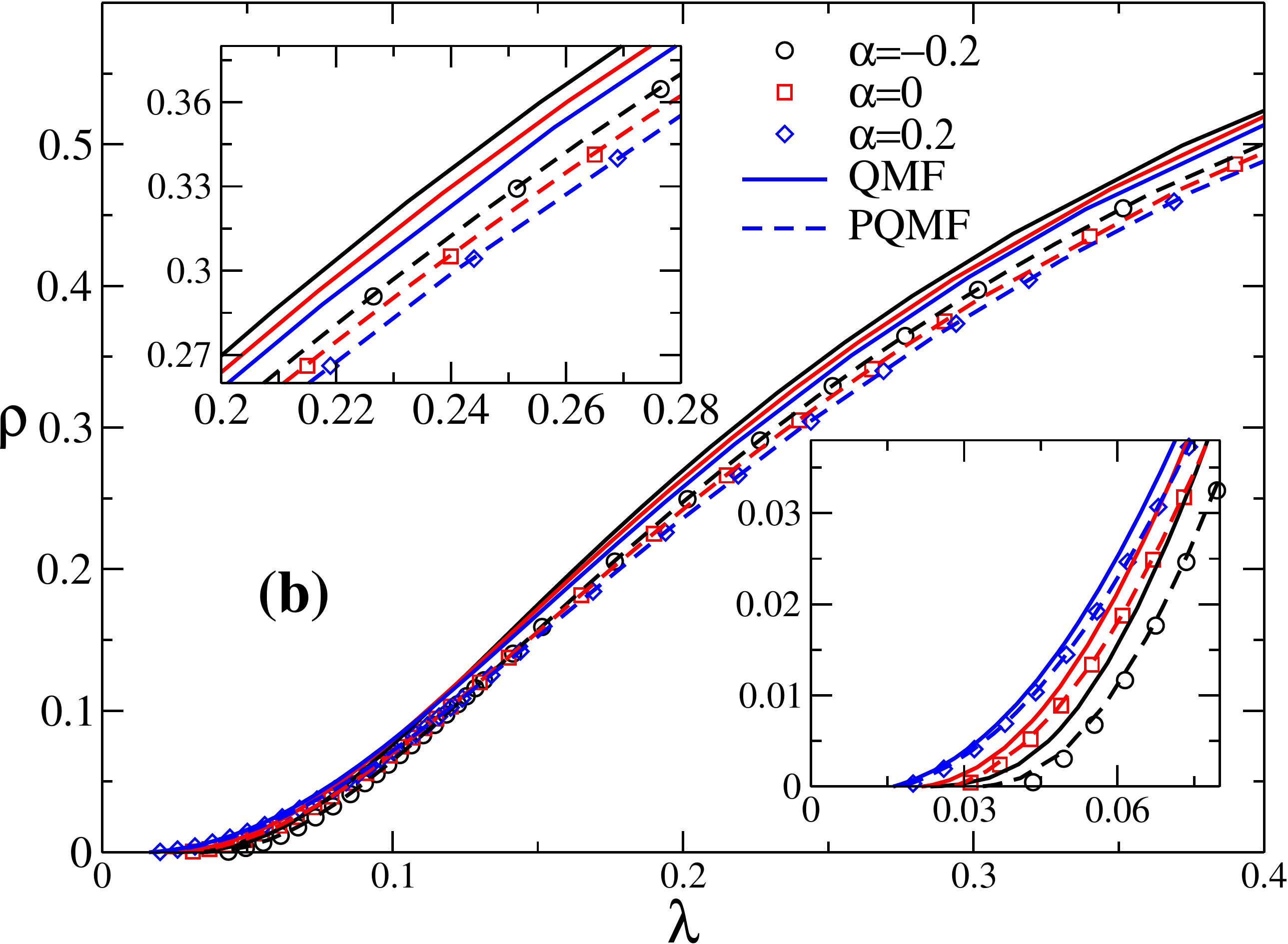}\\
	\includegraphics[width=0.8\linewidth]{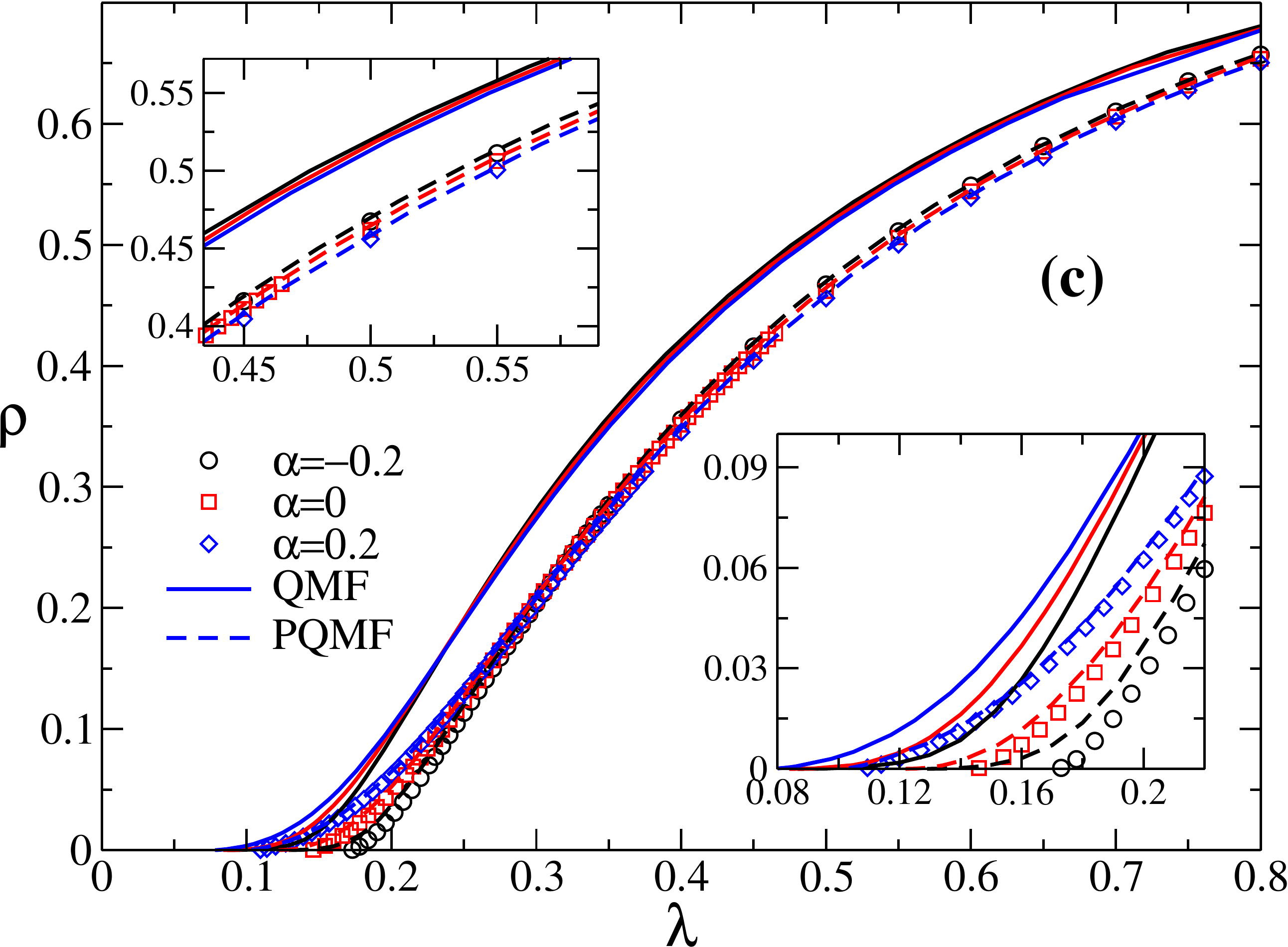}\\	
	\caption{Epidemic prevalence as a function of the infection rate for WPCM
		networks with $N=10^6$ vertices, degree exponent (a) $\gamma=2.3$, (b) 2.8 and
		(c) 3.5 and different levels of degree correlations. Symbols represent
		stochastic simulation while solid and dashed lines the numerical
		integration of the QMF and PQMF theories, respectively. Bottom and top insets
		show zoom of low and high prevalence, respectively.} \label{fig:rhoxlambda_g230_1e6_fmt}
\end{figure}
We investigated networks with distributions $P(k)\propto k^{-\gamma}$ and
$k=\kmin,\ldots,\kmax$ where $\kmin=3$. For $\gamma<3$ we used $\kmax=2\sqrt{N}$
that permits to build networks without degree correlation with the uncorrelated
configuration model (UCM)~\cite{Catanzaro2005}. The factor 2 helps to
accelerate the convergence to the asymptotic limit where  both $N\rightarrow\infty$
and $\kmax\rightarrow\infty$~\cite{Silva2019}. For $\gamma>3$, a rigid cutoff
given by $P(\kmax)N=1$~\cite{Dorogovtsev2008} was used to suppress multiple
(localized) transitions~\cite{Mata2015} and facilitating threshold
determination. Degree correlations were included using the benchmark model
proposed by Weber and Porto~\cite{Weber2007}, hereafter called Weber-Porto
configuration  model (WPCM); see Appendix~\ref{app:wp}. The dependence
$\kappa_{nn}(k)\propto k^\alpha$ was investigated where $\alpha<0$, $\alpha=0$ and
$\alpha>0$ correspond to disassortative, neutral, and assortative
correlations respectively.

Epidemic prevalence obtained from theories and stochastic simulations are
compared  in Fig.~\ref{fig:rhoxlambda_g230_1e6_fmt}. While QMF theory deviates
from simulation for regimes of high densities of infected vertices,  PQMF cannot
be distinguished from simulations in the presented scales. At low densities
regimes, QMF and PQMF agree very well and are indistinguishable from simulations
for $\gamma=2.3$, Fig.~\ref{fig:rhoxlambda_g230_1e6_fmt}(a), while larger
deviations of QMF can be seen for larger values of $\gamma$,
Figs~\ref{fig:rhoxlambda_g230_1e6_fmt}(b) and (c). The accuracy of the theories
at low prevalence is better for assortative and worse for disassortative
networks when compared with the neutral case. Another interesting dependence on the
assortativity can be observed in these curves. For low prevalence,  assortative
and disassortative networks possess, respectively, higher and lower densities if
compared with the uncorrelated networks. At high prevalences, the converse is
observed where disassortative networks present higher densities than the
assortative and neutral networks. The same behavior is observed for all values
of $\gamma$, indicating that it is related to the degree correlations. The
behavior at low densities can be explained in terms of reduced capacity to
transmit infection when hubs are surrounded by low degree vertices in the
disassortative case rather than being directly connected more likely in the
assortative case. We cannot provide simple arguments for  inverted dependence on
the assortativity degree at high densities, but it is very precisely reproduced
by the PQMF theory. We also simulated the SIS on larger networks with $N=10^7$
vertices and the level of accuracy of the mean-field theories is  similar.

\begin{figure}[tbh]
	\centering
	\includegraphics[width=0.8\linewidth]{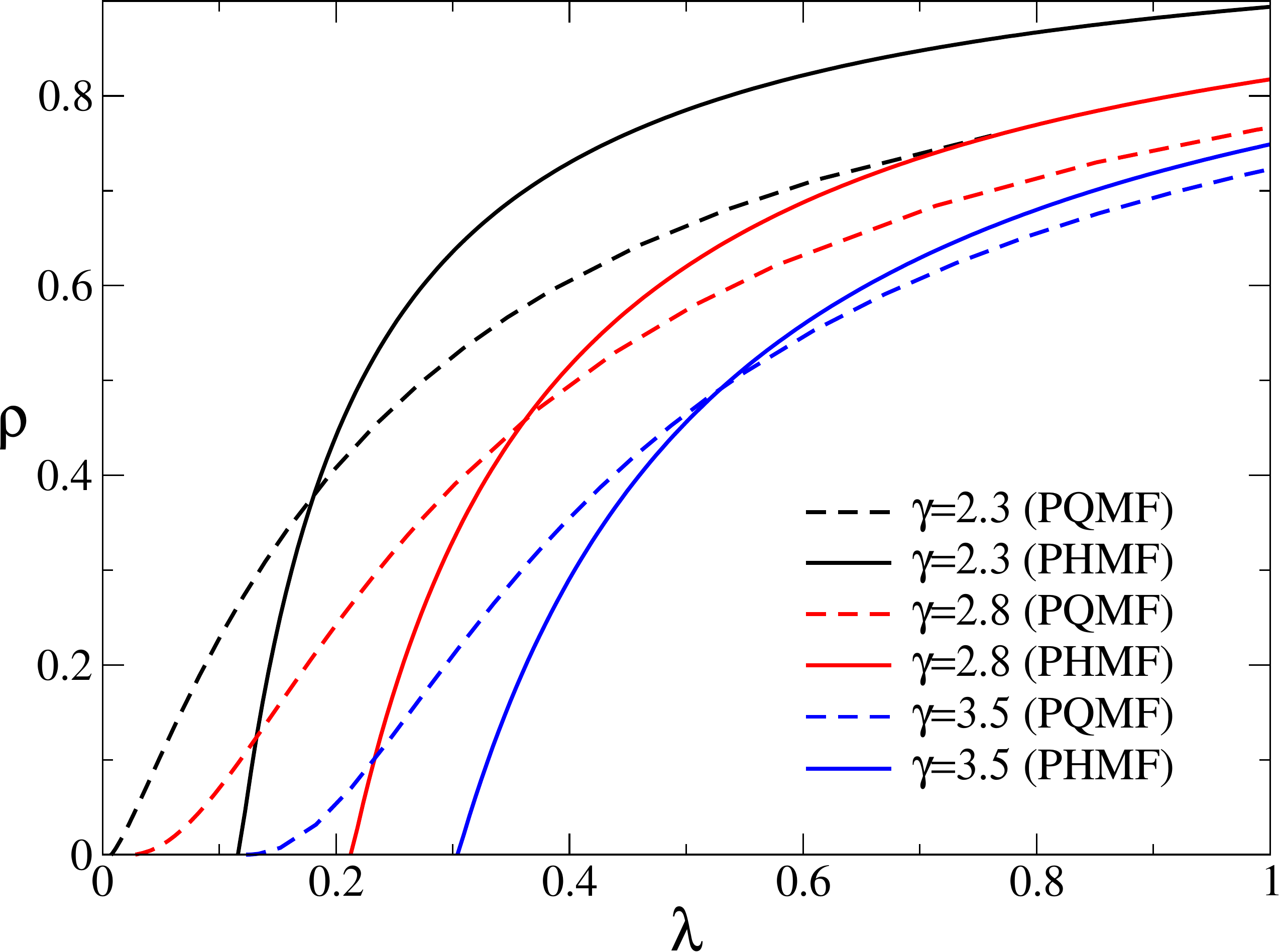}
	\caption{Comparison of pair quenched (dashed lines) and homogeneous (solid
		lines) mean-field theories for uncorrelated networks with different levels of
		heterogeneity. The network size is $N=10^6$.}
	\label{fig:rhoxlambda_1e6_PMF}
\end{figure}
We also investigate the role of the heterogeneity comparing the PQMF theory with
a pair homogeneous mean-field theory~(PHMF) using Eq.~\eqref{eq:phomo} with $m$
replaced by the average degree  $\av{k}$ of the
network~\cite{Juhasz2012,DeOliveira2019}. The density of infected vertices
obtained in both pairwise approaches are shown in
Fig.~\ref{fig:rhoxlambda_1e6_PMF}. Beyond the expected discrepancy for
describing the low prevalence regimes, since one theory predicts a finite while
the other a vanishing threshold, the regime of high epidemic prevalence is affected by inclusion of  heterogeneity. As one could expect, the more
heterogeneous networks present the larger discrepancies between homogeneous and
heterogeneous theories.

\subsection{Real networks}
\begin{figure*}[tbh]
	\centering
	\includegraphics[width=0.75\linewidth]{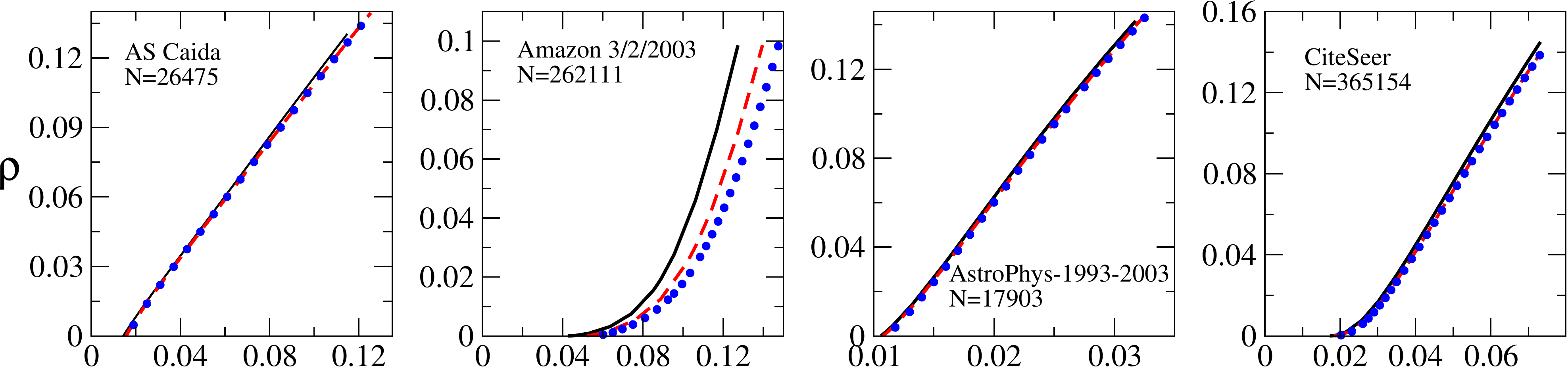}\\ \vspace{0.15cm}
	{\includegraphics[width=0.75\linewidth]{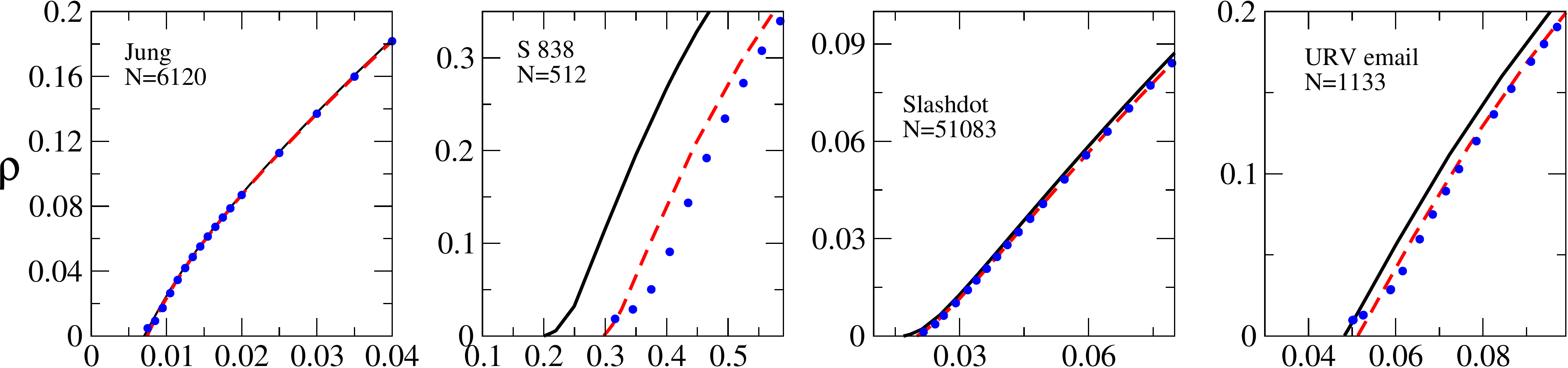}}\\ \vspace{0.15cm}
	\includegraphics[width=0.75\linewidth]{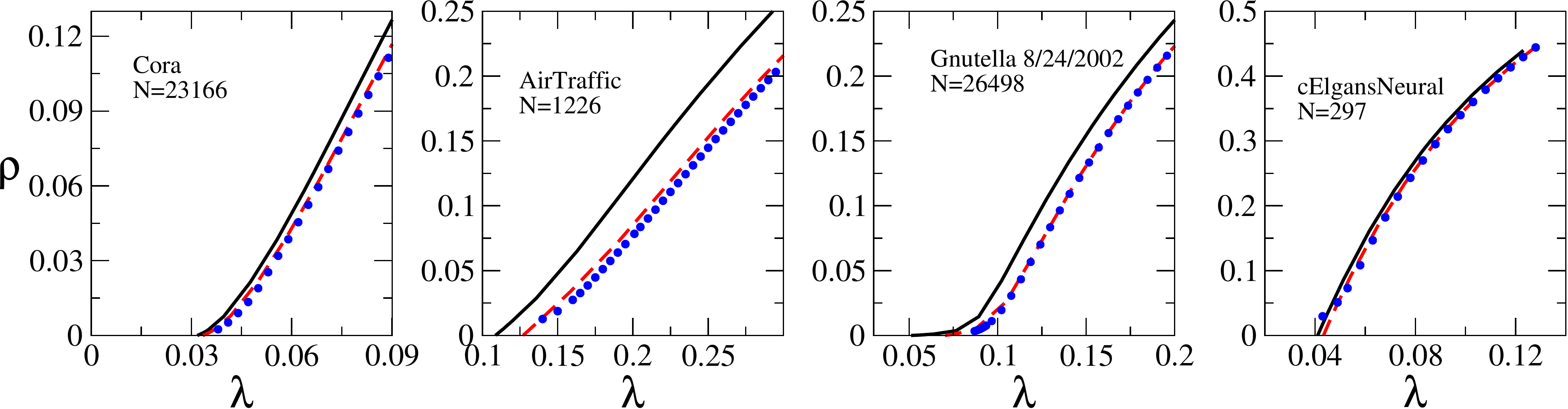}
	\caption{Epidemic prevalence on real networks. Symbols represent stochastic
		simulations while solid and dashed lines represent the numerical integration of
		the QMF and PQMF equations, respectively.  In each panel, the usual name and
		size of networks are given.}
	\label{fig:real}
\end{figure*}
Real networks usually present some degree of correlation and, in many case, the
patterns can be quite complex exhibiting both assortative and disassortative
correlations for distinct ranges of
degree~\cite{Barrat2008,barabasi2016network}. Therefore the comparison between
mean-field theories and simulations are  necessary in order to determine in
which extent the accuracy observed in synthetic networks holds in the real-world
counterparts. We selected some networks with different levels of heterogeneity,
sizes, and correlations recently used in the investigation of epidemic
processes~\cite{Silva2019,Castellano2017,Pastor-Satorras2018}. For detailed
information about the original references for all the networks see
Refs.~\cite{Radicchi2015,Radicchi2015b}.

Figure~\ref{fig:real} presents the prevalence as a function of the infection rate
for 12 real networks. We remark that data asymptotically close to the epidemic
threshold are known to mismatch simulations~\cite{Silva2019} and are beyond the
scope of the present work. In some cases, QMF and PQMF are indistinguishable from
each other and agree almost perfectly with simulations in the scale presented in
these figures. In other cases, QMF theory deviates considerably from
simulations while PQMF remains accurate. In order to quantify the differences we
define the relative deviation of densities obtained in simulations
($\rho_\text{sim}$) and  the QMF  theory ($\rho_\text{QMF}$) as
\begin{equation}
\eta_\text{QMF} = \frac{\int_{\lambda_1}^{\lambda_2}[\rho_\text{QMF}(\lambda)-\rho_\text{sim}(\lambda)]d\lambda}
{\int_{\lambda_1}^{\lambda_2}\rho_\text{sim}(\lambda) d\lambda},
\end{equation}
where  $\lambda_1$ and $\lambda_2$ are the initial and final infection rates in the simulations presented in Fig.~\ref{fig:real}, and an equivalent definition of $\eta_\text{PQMF}$ for
the PQMF theory. The relative deviations are given in Table~\ref{tab:real}. The
positivity of $\eta$ shows that the mean-field theories overestimate the density
obtained in simulations  as expected since dynamical correlations, pruned in
mean-field theories,  reduce the spreading capacity of the epidemic process.  As
can be seen, we have $\eta_\text{PQMF}\ll\eta_\text{QMF}$ such that PQMF is
always more precise than QMF.  Three networks present significant deviations of the
PQMF theory from the simulations, namely, for Amazon customer,  electronic
circuit S838, and air traffic networks with 24\%, 7\% and 6.6\% of deviation, respectively.

\subsection{Accuracy versus structural properties}

\begin{figure}[tbh]
	\centering
	\includegraphics[width=0.97\linewidth]{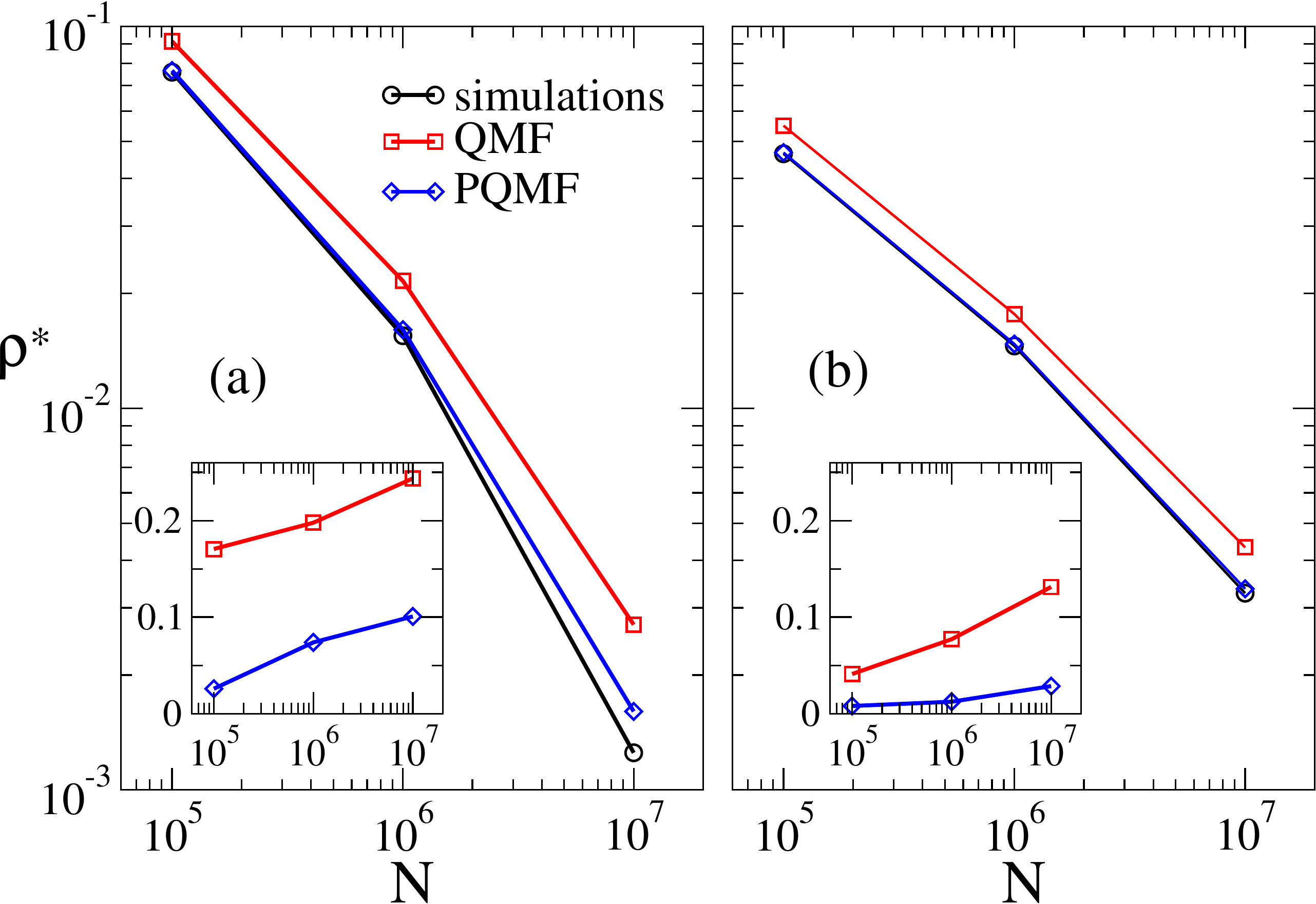}
	\caption{Finite size scaling of the steady-state density evaluated at
		$\lambda=2\lambda^\text{PQMF}_\text{c}$. WPCM networks with degree exponent
		$\gamma=2.8$ presenting (a) disassortative ($\alpha=-0.2$)  and (b) neutral
		($\alpha=0$) degree correlations  are considered. Insets show the corresponding
		IPR calculated for PEV of the corresponding mean-field theory. }
	\label{fig:fss}
\end{figure}

The gain of PQMF theory  with respect to QMF in real networks is expressive but
it still deviates from simulations in some cases, as shown in
Table~\ref{tab:real}. One central question is to determine when either QMF or
PQMF performance is satisfactory enough. Near to the transition point, when the
prevalence is very small, a relation between accuracy and localization of the PEV of
the weighted adjacency matrices obtained in the linearization has been
proposed~\cite{Silva2019}. This is justified by the fact that a leading
contribution for the probability that a vertex $i$ is infected in the mean-field
theories is proportional to the corresponding PEV of $A_{ij}$ or
$B_{ij}(\lambda_\text{c})$ for QMF and PQMF theories, respectively. For sake of
completeness of Ref.~\cite{Silva2019}, in which the accuracy on epidemic
threshold was discussed thoroughly, Fig.~\ref{fig:fss} shows the steady state
density calculated slightly above the epidemic threshold of the PQMF theory
against the network size.  The PQMF theory is much more  accurate than QMF but also
start to deviate from simulations as the networks size increases. In both cases
the accuracy is larger for networks with less localized PEV as quantified by the
inverse participation ratio (IPR)~\cite{Goltsev2012} defined as
	\begin{equation}
	Y^{(1)} = \sum_{i=1}^{N} \left[v_i^{(1)}\right]^4,
	\label{eq:IPR}
	\end{equation}
where ${v_i^{(1)}}$ is the normalized PEV defined in Sec.~\ref{sec:qmf} . As
larger as the IPR, more localized is the PEV. Insets of
Fig.~\ref{fig:fss} shows the IPR for both QMF and PQMF theories where we see
that the latter is much less localized than the former, but still increases
towards a finite value as the network size increases indicating  localization
asymptotically.

\begin{table}[hbt]
	\centering
	\begin{tabular*}{\linewidth}{l @{\extracolsep{\fill}} cccc}\hline\hline
		Network            & $\eta_\text{QMF}$ & $\eta_\text{PQMF}$ & $Y^{(1)}_\text{QMF}$ & $Y^{(1)}_\text{PQMF}$ \\\hline
		AS Caida           & 0.032 & 0.0022& 0.0240& 0.0139 \\ 
		Amazon 3/2/03      &0.75   & 0.24  & 0.106 & 0.0114\\
		AstroPhys 93-03    &0.035  & 0.013 & 0.0045& 0.0043\\
		CiteSeer           &0.067  & 0.010 & 0.0177& 0.0109\\
		Jung               &0.0079 & 0.0021& 0.0478& 0.0335\\
		S 838              &0.28   & 0.070 & 0.179 & 0.0340\\
		Slashdot           &0.032  & 0.0011& 0.144 & 0.0347\\
		URV email          &0.022  & 0.0028& 0.0096& 0.0087\\
		Cora               &0.152  & 0.037 & 0.0100& 0.0090\\
		Air Traffic        & 0.38  & 0.066 & 0.0191& 0.0154\\
		Gnutella 8/24/02   & 0.17  & 0.0066& 0.214 & 0.0800\\
		cElegans Neural    & 0.057 & 0.0078& 0.0189& 0.0175\\
		
		\hline\hline
	\end{tabular*}
	\caption{Relative deviations and inverse participation ratios for QMF and PQMF theories applied
		to real networks.}
	\label{tab:real}
\end{table}

However, the  nonperturbative theory accounts for the contributions of the complete basis of eigenvectors, whether it is of 
$A_{ij}$ or  $B_{ij}(\lambda_\text{c})$, and this comparison is not justifiable
anymore. Indeed, as shown in Table~\ref{tab:real}, the accuracy of the QMF theory can be
high even when the PEV is localized as, for example, in the case of the Slashdot
network. In other cases,  as  Air Traffic and Cora networks, the  PEV
localization corresponding to QMF and PQMF theories are similar but the
performance of the latter is much better. We performed a statistical analysis of the 
correlations between $\ln \eta$ and $\ln Y^{(1)}$ and found no statistically
significant $p$-values of $p_{_\text{QMF}}=0.29$ and $p_{_\text{PQMF}}=0.46$. It
is worth to note that the statistical analyses considering the linear data
present even smaller statistical significance.

We  also  checked (logarithm) statistical correlations of  $\eta$ with other
basic network metrics, namely, the heterogeneity coefficient
$\varepsilon=\av{k^2}/\av{k}$, the modularity coefficient
$Q$~\cite{newman2010networks}, the average  clustering coefficient $\av{c}$, and
average shortest distances $\av{\ell}$. We found statistical significance with
$p<0.02$ only with $\varepsilon$ and $\av{l}$. Actually, $\varepsilon$ and
$\av{l}$ are correlated since  more heterogeneous networks tend to have shortest
average distances due to the shortcuts introduced by
hubs~\cite{barabasi2016network}. The correlation between $\eta$ and $\av{\ell}$
actually is not very surprising since one intuitively expects that the shorter are
distances more the mean-field hypothesis of neglecting long-range correlations
becomes accurate. One interesting feature is that the approximation given by
Eq.~\eqref{eq:pair} in the PQMF theory discards the possibility  of
triangles~\cite{Mata2013}, in which  the neighbors $i$ and $l$ of $j$ are also
connected. So, one could expect a worse performance in networks with high
clustering coefficient but no statistical correlation with this metrics was
found ($p_{_\text{PQMF}}=0.51$). In summary, we could not infer which structural
properties rule the accuracy of the mean-field approaches in the regimes of high
prevalences.

\section{Conclusions}
\label{sec:conclu}

Theoretical understanding of dynamical processes on networks constitutes a
powerful tool for protection against threats such as disease dissemination,
misinformation propagations, transportation infrastructure overload, among many
other examples. Reliable  theoretical approximations  are usually required to consider the heterogeneous structure of the
contact networks and the dynamical correlations, in which the states of
neighboring individuals are statistically correlated. These features are
explicitly included in  the PQMF theory~\cite{Mata2013}. However, this theory
has been applied mainly to analyze the behavior of epidemic processes in the
neighborhood of the transition from an endemic to a disease-free state trough
perturbative analyses where the epidemic prevalence is very small. In this work,
we contribute to fill this gap performing a detailed nonperturbative numerical
analysis of the SIS model on synthetic and real networks within a wide range of
heterogeneities and assortativities.

For synthetic networks generated with the Weber-Porto~\cite{Weber2007}
configuration model, we report that the PQMF theory predicts with great accuracy the regime of high prevalence observed in stochastic simulations in networks with power-law degree distributions for all values of degree exponents
investigated ($\gamma=2.3$, 2.8, and 3.5) and degree correlations (disassortative, neutral and assortative). In the case of large $\gamma>2.5$, where hubs tend
to be separated apart as the network size increases, we observed that the PQMF
theory  significantly outperforms the simpler QMF theory where heterogeneity is
fully considered but dynamical correlations are neglected, being the discrepancy
between theories larger for large $\gamma$. The high accuracy of the PQMF theory
at high prevalences contrasts with its bad performance for asymptotically low
densities where the theory is known to deviate from exactly known critical
behavior~\cite{Mountford2013} where $\rho\sim \lambda^{\beta}$ with $\beta>1$ while
mean-field exponent is $\beta_\text{MF}=1$~\cite{Silva2019}. We argue, however,
that this mismatch is constrained to a region very close to
$\lambda=\lambda_\text{c}\rightarrow 0^+$ such that the regime of not too low
density can still be accurately described by the PQMF theory.

In a set of real networks, where much more complex structures and correlations
can be present, we observed that PQMF always outperforms  (sometimes very
significantly) the QMF theory, but may still presents non negligible deviations
from simulations in some cases; see Table~\ref{tab:real}. Differently from the
low prevalence regime where the accuracy of mean-field theories is correlated
with spectral properties of Jacobian matrices, only trivial
statistical correlations with simple network metrics could be identified and the
problem of predicting when the nonperturbative analysis is sufficiently accurate
given certain network properties remains open.

Finally, we expect that our work will stimulate the application of
nonperturbative approaches through the numerical integration of continuous-time
equations to address other fundamental problems of dynamical processes on networks.

\appendix

\section{Weber-Porto configuration model}
\label{app:wp}

The WPCM networks are generated as follows. The degree of each vertex is drawn
according to the degree distribution $P(k)$ such that each node has $k$
unconnected half-edges. Two half-edges are chosen and connected with
probability
\begin{equation}
P_\text{link}(q',q)=\frac{f(q',q)}{f_\text{max}},
\label{eq:plink}
\end{equation}
where $q$ and $q'$ are the respective degrees of the chosen vertices and
$f_\text{max}$  is the maximum value of
\begin{equation}
f(q,q')=1+\frac{(\knn(q)-\ave{k})(\knn(q')-\ave{k})}{
	\ave{k\knn}-\ave{k}^{2}},
\label{eq:fq_qp}
\end{equation}
where $\ave{A(k)}=\sum_kA(k)P_\text{e}(k)$ where $P_\text{e}(k) =kP(k)/\av{k}$
is the probability that an edge ends on a vertex of degree $k$.  Self and
multiple connections are forbidden. In the absence of degree correlations, we
have $\knn=\ave{k}$, implying $f(q,q')=1$ and $P_\text{link}=1$.  See
Ref.~\cite{Weber2007} for more details.

\section{Quasi-stationary method}
\label{app:qs}

We applied the standard QS method~\cite{Sander2016,DeOliveira2005,Cota2017}
where the dynamics returns to a previously visited active configuration with at
least one infected vertex every time the system falls into the absorbing state
where all vertices are simultaneously susceptible. The method is implemented by
building and constantly updating a list with $M=100$ active configurations.
Every time the systems falls into the absorbing state one of the $M$
configurations is chosen with equal chance to replace the absorbing state. The
list is updated with probability $10^{-2}$ by unit of time and the update
consists of replacing a randomly selected configuration of the list by the
present state of the dynamics. The QS averages are computed  during an averaging
time varying from $t_\text{av}=10^5~\mu^{-1}$ to $10^6~\mu^{-1}$ after a 
relaxation time $t_\text{rlx}=10^5$, the longer times for the lower densities
where fluctuations are more relevant.

\section{Stochastic simulation of the SIS model}
\label{app:algo}

Simulations of SIS model were performed using the optimized Gillespie algorithm
described in~\cite{Cota2017}. The number of infected vertices $\Ninf$ and the 
total number of edges emanating from them $\NSI$ are computed and kept updated
along the simulations. In each time step, with probability 
\begin{equation}
q=\frac{\mu \Ninf}{\mu\Ninf+\lambda\NSI}
\end{equation}
one infected vertex is chosen with equal chance and healed.
With the complementary probability
\begin{equation}
1-q=\frac{\lambda\NSI}{\mu\Ninf+\lambda\NSI},
\end{equation}
one infected vertex $i$ is chosen with probability proportional to its degree. 
One of nearest-neighbors of $i$, represented by $j$, is chosen with equal chance.
If $j$ is susceptible, it becomes infected and, otherwise, no change of state is
implemented.  The time is incremented by 
\begin{equation}
\delta t  = \frac{-\ln u}{\mu\Ninf+\lambda\NSI}
\end{equation}
where $u$ is a pseudo random number  uniformly distributed in the interval $(0,1)$.

\begin{acknowledgments}
	This work was partially supported by the Brazilian agencies CNPq and FAPEMIG.
	This study was financed in part by the Coordenação de Aperfeiçoamento de
	Pessoal de Nível Superior - Brasil (CAPES) - Finance Code 001.
\end{acknowledgments}

\end{document}